\documentclass[twocolumn,showpacs,preprintnumbers,amsmath,amssymb]{revtex4}

\usepackage{graphicx}
\usepackage{dcolumn}
\usepackage{bm}
\begin{document}

\preprint{}

\title{Non-perturbative solutions in the\\ electro-weak theory and the Higgs\\ boson mass}

\author{Boris A. Arbuzov }

\affiliation{Skobeltsyn Institute for Nuclear Physics of Moscow State University\\ Leninskie gory 1, 119992 Moscow, Russia}%
\email{arbuzov@theory.sinp.msu.ru}
\author{Ivan V. Zaitsev}
\affiliation{Skobeltsyn Institute of Nuclear Physics of
Moscow State University\\ Leninskie gory 1,119992 Moscow, Russia}
\date{\today}
\begin{abstract}
We apply Bogoliubov compensation principle to the gauge electro-weak interaction
to demonstrate a  spontaneous generation of anomalous three-boson gauge invariant
effective interaction. The non-trivial solution of compensation equations uniquely defines the form-factor of the anomalous interaction and parameters of the theory including value of gauge electro-weak coupling $g(M_W^2)\simeq 0.62$ in satisfactory agreement to the experimental value.
A possibility of spontaneous generation of effective four-fermion interaction of heavy quarks is also demonstrated.
This interaction defines an equation for a scalar bound state of heavy quarks which serve as a substitute for the elementary scalar Higgs doublet. As a result we obtain two possible solutions for the observable Higgs boson mass: $M_1\,=\,(306\,\pm\,14)GeV$ and $M_2\,=\,(253\,\pm\,10)\,GeV$. Values for other parameters e.g. anomalous three-boson coupling $\,G$, Higgs boson width $\Gamma_H$, its branching ratios and estimates for cross-section of Higgs production at LHC are also obtained.
\end{abstract}

\maketitle

In previous works~\cite{Arb04, Arb05, AVZ, Arb07, AVZ2, Arb09}
N.N. Bogoliubov compensation principle~\cite{Bog1, Bog2, Bog} was applied to studies of spontaneous generation of effective non-local interactions in renormalizable
gauge theories.  Spontaneous generation of Nambu -- Jona-Lasinio like
interaction was studied in works~\cite{Arb05, AVZ, AVZ2} and the description of low-energy
hadron physics in terms of initial QCD parameters turns to be quite successful including values
of parameters: $m_\pi,\, f_\pi,\, m_\sigma,\, 
\Gamma_\sigma,\,  M_\rho,\,  \Gamma_\rho,\,  M_{a_1},\,  \Gamma_{a_1},\, <\bar q q>$.

In work~\cite{Arb09}  the approach was applied to the electro-weak interaction and a possibility of spontaneous generation of anomalous three-boson interaction of the form
\begin{equation}
-\,\frac{G}{3!}\cdot\,\epsilon_{abc}\,W_{\mu\nu}^a\,W_{\nu\rho}^b\,W_{\rho\mu}^c\,;
\label{FFF}
\end{equation}
was studied. In the present work we continue investigation
of the electro-weak theory using other approximation scheme,
which will be formulated in what follows.

The main principle of the approach is to check if an effective interaction
could be generated in a chosen variant of a renormalizable theory. In view
of this one performs "add and subtract" procedure for the effective
interaction with a form-factor. Then one assumes the presence of the
effective interaction in the interaction Lagrangian and the same term with
the opposite sign is assigned to the newly defined free Lagrangian. This
transformation of the initial Lagrangian is evidently identical. However
such free Lagrangian contains completely improper term, corresponding to
the effective interaction of the opposite sign. Then one
has to formulate a compensation equation, which guarantees that this new
free Lagrangian is a genuine free one, that is effects of the
uncommon term sum up to zero. Provided a non-trivial solution of this
equation exists, one can state the generation of the effective
interaction to be possible. Now we apply this procedure to our problem.

In the present work we start with studying of possibility of generation of interaction~(\ref{FFF}).

\section{Compensation equation for anomalous three-boson interaction}\label{alpha}

We start with EW Lagrangian with $3$ lepton $l_k$ and colour quark $q_k$ spinor doublets
with gauge group $SU(2)$. That is we restrict the gauge sector to
triplet of $W^a_\mu$ only. Thus we consider $U(1)$ abelian gauge field $B$ to be decoupled, that means approximation $\sin^2\theta_W \ll 1$.
\begin{eqnarray}
& & L = \sum_{k=1}^3 \frac{\imath}{2}
\Bigl(\bar l_k\gamma_
\mu\partial_\mu l_k\,-\partial_\mu\bar l_k
\gamma_\mu l_k\,+ \nonumber\\
& &\bar q_k\gamma_
\mu\partial_\mu q_k\,-\partial_\mu\bar q_k
\gamma_\mu q_k\,\Bigr)\,-\,\frac{1}{4}\,W_{\mu\nu}^a W_{\mu\nu}^a\,+\nonumber\\
& &\frac{g}{2} \sum_{k=1}^3\Bigl(\bar l_{kL}\gamma_\mu \tau^a l_{kL} + \bar q_{kL}\gamma_\mu \tau^a q_{kL}\Bigr) W_\mu^a\,;\label{initial}\\
& & W_{\mu\nu}^a =
\partial_\mu W_\nu^a - \partial_\nu W_\mu^a + g\,\epsilon_{abc}W_\mu^b W_\nu^c\,;
\nonumber\\
& &\psi_{kL}=\frac{1+\gamma_5}{2}\psi_k\,.\nonumber
\end{eqnarray}
where we use the standard notations. Let us draw attention to the absence of the Higgs sector of the theory. We shall 
obtain Higgs scalars to be bound states of heavy quarks. The effective interaction of Higgs fields will be obtained as well.
 
In accordance to the Bogoliubov approach~\cite{Bog1, Bog2, Bog} in application to QFT~\cite{Arb04} we look for
a non-trivial solution of a
compensation equation, which is formulated on the basis
of the Bogoliubov procedure {\bf add -- subtract}. Namely
let us write down the initial expression~(\ref{initial})
in the following form
\begin{eqnarray}
& &L\,=\,L_0\,+\,L_{int}\,;\nonumber\\
& &L_0\,=\,\sum_{k=1}^3 \frac{\imath}{2}
\Bigl(\bar l_k\gamma_
\mu\partial_\mu l_k\,-\partial_\mu\bar l_k
\gamma_\mu l_k\,+ \nonumber\\\label{L0}
& &\bar q_k\gamma_
\mu\partial_\mu q_k\,-\partial_\mu\bar q_k
\gamma_\mu q_k\,\Bigr)\,-\,\frac{1}{4}\,W_{\mu\nu}^a W_{\mu\nu}^a\,+\nonumber\\
& &\frac{G}{3!}\cdot\,\epsilon_{abc}\,W_{\mu\nu}^a\,W_{\nu\rho}^b\,W_{\rho\mu}^c\,;
\\
& &L_{int} = \frac{g}{2} \sum_{k=1}^3\Bigl(\bar l_{kL}\gamma_\mu \tau^a l_{kL} + \bar q_{kL}\gamma_\mu \tau^a q_{kL}\Bigr) W_\mu^a - \nonumber\\
& &-\,\frac{G}{3!}\cdot\,\epsilon_{abc}\,
W_{\mu\nu}^a\,W_{\nu\rho}^b\,W_{\rho\mu}^c\,.\label{Lint}
\end{eqnarray}
Here isotopic summation is
performed inside of each spinor 
bi-linear combination, and notation
 $-\,\frac{G}{3!}\cdot \,\epsilon_{abc}\,
W_{\mu\nu}^a\,W_{\nu\rho}^b\,W_{\rho\mu}^c$ means corresponding
non-local vertex in the momentum space
\begin{eqnarray}
& &(2\pi)^4\,G\,\,\epsilon_{abc}\,(g_{\mu\nu} (q_\rho pk - p_\rho qk)+ g_{\nu\rho}
(k_\mu pq - q_\mu pk)+\nonumber\\
& &+g_{\rho\mu} (p_\nu qk - k_\nu pq)+\,q_\mu k_\nu p_\rho - k_\mu p_\nu q_\rho)\times\nonumber\\
& &\times F(p,q,k)\,
\delta(p+q+k)\,+...;\label{vertex}
\end{eqnarray}
where $F(p,q,k)$ is a form-factor and
$p,\mu, a;\;q,\nu, b;\;k,\rho, c$ are respectfully incoming momenta,
Lorentz indices and weak isotopic indices
of $W$-bosons. We mean also that there are present four-boson, five-boson and
six-boson vertices according to expression for $W_{\mu\nu}^a$
(\ref{initial}).

Effective interaction~(\ref{FFF}) is
usually called {\bf anomalous three-boson interaction} and it is considered for long time on phenomenological grounds~\cite{Hag}. Note, that the first attempt to obtain the anomalous three-boson interaction in the framework of Bogoliubov approach was done in work~\cite{Arb92}. Our interaction constant $G$ is connected with
conventional definitions in the following way
\begin{equation}
G\,=\,-\,\frac{g\,\lambda}{M_W^2}\,.\label{Glam}
\end{equation}
The current limitations for parameter $\lambda$ read~\cite{EW}
\begin{eqnarray}
& &\lambda = -\,0.016^{+0.021}_{-0.023};\label{lambda1}\\
& & -\,0.059< \lambda < 0.026\,(95\%\,C.L.)\,.\nonumber
\end{eqnarray}
Due to our approximation $\sin^2\theta_W\,
\ll\,1$ we use the same $M_W$ for both charged $W^\pm$ and neutral $W^0$ bosons and assume no difference in anomalous interaction for $Z$ and $\gamma$, i.e.
$\lambda_Z\,=\,\lambda_\gamma\,=\,\lambda$.

Let us consider  expression~
(\ref{L0}) as the new {\bf free} Lagrangian $L_0$,
whereas expression~(\ref{Lint}) as the new
{\bf interaction} Lagrangian $L_{int}$. It is important to note, that we
put into the new {\bf free} Lagrangian the full quadratic in $W$ term including
boson self-interaction, because we prefer to maintain gauge invariance of the approximation being used. Indeed, we shall use both quartic term from the last term
in~(\ref{L0}) and triple one from the last but one term of~(\ref{L0}).
Then compensation conditions (see for details~\cite{Arb04}) will
consist in demand of full connected three-gluon vertices of the structure~(\ref{vertex}),
following from Lagrangian $L_0$, to be zero. This demand
gives a non-linear equation for form-factor $F$.

Such equations according to terminology of works
~\cite{Bog1, Bog2, Bog} are called {\bf compensation equations}.
In a study of these equations it is always evident the
existence of a perturbative trivial solution (in our case
$G = 0$), but, in general, a non-perturbative
non-trivial solution may also exist. Just the quest of
a non-trivial solution inspires the main interest in such
problems. Let us remind, that non-trivial solutions give proper explanations for super-fluidity and superconductivity effects (see again~\cite{Bog1, Bog2, Bog}). One can not succeed in finding an exact
non-trivial solution in a realistic theory, therefore
the goal of a study is a quest of an adequate
approach, the first non-perturbative approximation of
which describes the main features of the problem.
Improvement of a precision of results is to be achieved
by corrections to the initial first approximation.

Thus our task is to formulate the first approximation.
Here the experience acquired in the course of performing
works~\cite{Arb04, Arb05, AVZ, Arb07} could be helpful. Now in view of
obtaining the first approximation we would make the following
assumptions.\\
1) In compensation equation we restrict ourselves by
terms with loop numbers 0, 1.\\
2) We reduce thus obtained non-linear compensation equation to a linear
integral equation. It means that in loop terms only one vertex
contains the form-factor, being defined above, while
other vertices are considered to be point-like. In
diagram form equation for form-factor $F$ is presented
in Fig. 1. Here four-leg vertex correspond to interaction of four
gluons due to our effective three-field interaction. In our approximation we
take here point-like vertex with interaction constant proportional
to $g\,G$.\\
3) We integrate by angular variables of the 4-dimensional Euclidean
space. The necessary rules are presented in paper~\cite{Arb05}.

At first let us present the expression for four-boson vertex
\begin{widetext}
\begin{eqnarray}
& &V(p,m,\lambda;\,q,n,\sigma;\,k,r,\tau;\,l,s,\pi) = (2\,\pi)^4\,g G \biggl(\epsilon^{amn}
\epsilon^{ars}\Bigl(U(k,l;\sigma,\tau,\pi,\lambda)-U(k,l;\lambda,\tau,\pi,\sigma)-U(l,k;\sigma,\pi,\tau,\lambda)+\nonumber\\
& &
U(l,k;\lambda,\pi,\tau,\sigma)+U(p,q;\pi,\lambda,\sigma,\tau)-U(p,q;\tau,\lambda,\sigma,\pi)-U(q,p;\pi,\sigma,\lambda,\tau)
+U(q,p;\tau,\sigma,\lambda,\pi)\Bigr)-\nonumber\\
& &\epsilon^{arn}\,
\epsilon^{ams}\Bigl(U(p,l;\sigma,\lambda,\pi,\tau)-U(l,p;\sigma,\pi,\lambda,\tau)
-U(p,l;\tau,\lambda,\pi,\sigma)+
U(l,p;\tau,\pi,\lambda,\sigma)+U(k,q;\pi,\tau,\sigma,\lambda)-\label{UU}\\
& &U(q,k;\pi,\sigma,\tau,\lambda)
-U(k,q;\lambda,\tau,\sigma,\pi)
+U(q,k;\lambda,\sigma,\tau,\pi)\Bigr)+\epsilon^{asn}\,
\epsilon^{amr}\Bigl(U(k,p;\sigma,\lambda,\tau,\pi)-U(p,k;\sigma,\tau,\lambda,\pi)
+\nonumber\\
& &U(p,k;\pi,\tau,\lambda,\sigma)-U(k,p;\pi,\lambda,\tau,\sigma)-U(l,q;\lambda,\pi,\sigma,\tau)
+U(l,q;\tau,\pi,\sigma,\lambda)
-U(q,l;\tau,\sigma,\pi,\lambda)+U(q,l;\lambda,\sigma,\pi,\tau)\Bigr)\biggr)\,;\nonumber\\
& &U(k,l;\sigma,\tau,\pi,\lambda)=k_\sigma\,l_\tau\,g_{\pi\lambda}-k_\sigma\,l_\lambda\,g_{\pi\tau}+k_\pi\,l_\lambda\,g_{\sigma\tau}-(kl)g_{\sigma\tau}g_{\pi\lambda}\,.\nonumber
\end{eqnarray}
\end{widetext}
Here triad $p,\,m,\,\lambda$ {\it etc} means correspondingly incoming momentum, isotopic
index, Lorentz index of a boson.

Let us formulate compensation equations in this
approximation.
For {\bf free} Lagrangian $L_0$ full connected
three-gluon vertices with Lorentz structure~(\ref{vertex}) are to vanish. One can succeed in
obtaining analytic solutions for the following set
of momentum variables (see Fig. 1): left-hand legs
have momenta  $p$ and $-p$, and a right-hand leg
has zero momenta.
However in our approximation we need form-factor $F$ also
for non-zero values of this momentum. We look for a solution
with the following simple dependence on all three variables
\begin{equation}
F(p_1,\,p_2,\,p_3)\,=\,F(\frac{p_1^2\,+\,p_2^2\,+\,p_3^2}{2})\,;\label{123}
\end{equation}
Really, expression~(\ref{123}) is symmetric and it turns to $F(x)$
for $p_3=0,\,p_1^2\,=\,p_2^2\,=\,x$. We consider the representation~(\ref{123})
to be the first approximation and we plan to take into account the
corresponding correction in forthcoming studies.

We shall look for non-trivial solutions of compensation equations. Such solutions are also evidently non-perturbative. We would emphasize, that the existence of a non-trivial solution could be achived only provided some strict relations between parameters of the problem being fulfilled. 

Now according to the rules being stated above we
obtain the following equation for form-factor $F(x)$
\begin{widetext}
\begin{eqnarray}
& &F(x) = - \frac{G^2 N}{64 \pi^2}\Biggl(\int_0^Y F(y) y dy -
\frac{1}{12 x^2} \int_0^x F(y) y^3 dy 
+ \frac{1}{6 x} \int_0^x F(y) 
y^2 dy + \frac{x}{6} \int_x^Y F(y) dy - \frac{x^2}{12} 
\int_x^Y \frac{F(y)}{y} dy \Biggr) + \nonumber\\
& &\frac{G g N}{8 \pi^2}\Biggl( 
\int_0^Y\frac{F(y)}{2} dy + \int_{3 x/4}^{x}\,\frac{(3 x-  4 y)^2 (2 y -3 x)}{3 x^2 (x-2 y)}F(y) 
dy + \int_{x}^Y \frac{(5 x- 6 y)}{3 (x-2 y)}F(y)  dy +\int_x^Y \frac{3(x^2-2 y^2)}{32 (2 y-x)^2} F(y) dy+\nonumber\\
& &\int_{3 x/4}^{x} \frac{3(4 y-3 x)^2(x^2-4 x y+2 y^2)}{32 x^2(2 y-x)^2} F(y) dy 
+ \int_0^x\frac{5 y^2-12 x y}{64 x^2} F(y) dy + \int_x^Y 
\frac{3 x^2- 4 x y - 6 y^2}{64 y^2} F(y) dy\Biggr)\,.\label{eqF}
\end{eqnarray}
\end{widetext}
Here $x = p^2$ and $y = q^2$, where $q$ is an integration momentum, $N=2$. The last four terms in brackets represent diagrams with one usual gauge vertex (see three last
diagrams at Fig. 1). We introduce here
an effective cut-off $Y$, which bounds a "low-momentum" region where
our non-perturbative effects act
and consider the equation at interval $[0,\, Y]$ under condition
\begin{equation}
F(Y)\,=\,0\,. \label{Y0}
\end{equation}
 
We shall solve equation~(\ref{eqF}) by iterations. That is we
expand its terms being proportional to $g$ in powers of $x$ and
take at first only constant term. Thus we have
\begin{eqnarray}
& &F_0(x) = \nonumber\\
& &- \frac{G^2 N}{64 \pi^2}\Biggl(\int_0^Y F_0(y)y dy - \frac{1}{12 x^2} \int_0^x F_0(y) y^3 dy
+\nonumber\\
& &\frac{1}{6 x} \int_0^x F_0(y)
y^2 dy +\frac{x}{6} \int_x^Y F_0(y) dy -\label{eqF0}\\
& &\frac{x^2}{12} \int_x^Y \frac{F_0(y)}{y} dy \Biggr) + \frac{87 G g N}{512 \pi^2} \int_0^Y F_0(y) dy\,.\nonumber
\end{eqnarray}
Expression~(\ref{eqF0}) provides an equation of the type which were
studied in papers~\cite{Arb04, Arb05, AVZ, Arb07},
where the way of obtaining
solutions of equations analogous to (\ref{eqF0}) are described.
Indeed, by successive differentiation of Eq.(\ref{eqF0}) we come to
Meijer differential equation~\cite{be}
\begin{eqnarray}
& &\biggl(x \frac{d}{dx} + 2\biggr)\biggl(x \frac{d}{dx} + 1\biggr)\biggl(x \frac{d}{dx} - 1\biggr)\biggl(x \frac{d}{dx} - 2\biggr)F_0(x) +\nonumber\\
& &\frac{G^2 N x^2}{64 \pi^2} F_0(x) = 4 \Biggl(- \frac{G^2 N}{64 \pi^2} \int_0^Y F_0(y)
y dy +\label{difur}\\
& &\frac{87 G g N}{512 \pi^2} \int_0^Y F_0(y) dy
\Biggr);\nonumber
\end{eqnarray}
which solution looks like
\begin{eqnarray}
& &F_0(z) = \,C_1\,G_{04}^{10}\Bigl( z\,|1/2,\,1,\,-1/2,\,-1\Bigr) +\nonumber\\
& & C_2\,G_{04}^{10}\Bigl( z\,|1,\,1/2,\,-1/2,\,-1\Bigr)\,- \label{solution}\\
& & \frac{G^2 N}{128 \pi^2} G_{15}^{31}\Bigl( z |^0_{1, 1/2, 0, -1/2, -1}\Bigr) \int_0^Y \Biggl(y - \frac{87 g}{8 G}\Biggr)F_0(y)dy ;\nonumber\\
& & z\,=\,\frac{G^2\,N\,x^2}{1024\,\pi^2}\,;\nonumber
\end{eqnarray}
where
$$
G_{qp}^{nm}\Bigl( z\,|^{a_1,..., a_q}_{b_1,..., b_p}\Bigr)\,;
$$
is a Meijer function~\cite{be}. In case $q=0$ we write only indices $b_i$ in one
line. Constants $C_1,\,C_2$ are defined by the following boundary conditions
\begin{eqnarray}
& &\Bigl[2 z^2 \frac{d^3 F_0(z)}{dz^3} + 9 z \frac{d^2 F_0(z)}{dz^2} +
\frac{d F_0(z)}{dz}\Bigr]_{z = z_0} = 0\,;\nonumber\\
& &\Bigl[2 z^2 \frac{d^2 F_0(z)}{dz^2} + 5 z \frac{d F_0(z)}{dz} +
F_0(z) \Bigr]_{z = z_0} = 0\,;\label{bc}\\
& &
\quad z_0\,=\,\frac{G^2 N\,Y^2}{1024\,\pi^2}\,.\nonumber
\end{eqnarray}

Conditions~(\ref{Y0}, \ref{bc}) defines set of
parameters
\begin{equation}
z_0\,=\,\infty\,; \quad C_1\,=\,0\,
; \quad C_2\,=\,0\,.\label{z0C}
\end{equation}
The normalization condition for form-factor $F(0)=1$ here is the following
\begin{equation}
- \frac{G^2 N}{64 \pi^2} \int_0^\infty F_0(y)
 y dy + \frac{87 G g N}{512 \pi^2} \int_0^\infty F_0(y) dy = 1\,.
\label{norm}
\end{equation}
However the first integral in (\ref{norm}) diverges due to asymptotics
$$
G_{15}^{31}\Bigl( z\,|^0_{1,\,1/2,\,0,\,-1/2,\,-1}\Bigr)\,\to\,
\frac{1}{2\,z}\,, \quad z\,\to\,\infty\,;
$$
and we have no consistent solution. In view of this we consider the next
approximation. We substitute solution (\ref{solution}) with account of~(\ref{norm}) into terms of Eq.~(\ref{eqF}) containing gauge constant $g$ and
calculate terms being  proportional to $\sqrt{z}$. Now we have bearing in mind the normalization condition
\begin{eqnarray}
& &F(z) = 1 + \frac{85 g \sqrt{N z}}{96 \pi}\Biggl(\frac{1}{2}G_{15}^{31} \Bigl( z_0|^0_{0, 0, 1/2, -1, -1/2}\Bigr)
+\nonumber\\
& &\ln z + 4 \gamma + 4 \ln 2   - \frac{3160}{357}\Biggr) +
\frac{2}{3\,z} \int_0^z F(t)\,t\, dt + \nonumber\\
& &\frac{4}{3\,\sqrt{z}} \int_0^z F(t)
\sqrt{t}\, dt - \frac{4\,\sqrt{z}}{3} \int_z^{z_0} F(t) \frac{dt}{\sqrt{t}}\,+
\label{eqFg}\\
& &+\,\frac{2\,z}{3}\,\int_z^{z_0}\,F(t)\,\frac{dt}{t}\,;\nonumber
\end{eqnarray}
where $\gamma$ is the Euler constant. We look for solution of (\ref{eqFg})
in the form
\begin{eqnarray}
& &F(z)\,=\,\frac{1}{2}\,G_{15}^{31}\Bigl( z\,|^0_{1,\,1/2,\,0,\,-1/2,\,-1}
\Bigr) -\nonumber\\
& &\frac{85\,g \sqrt{N}}{128\,\pi}\,G_{15}^{31}\Bigl( z\,|^{1/2}_{1,\,1/2,
\,1/2,\,-1/2,\,-1}\Bigr)\,+\nonumber\\
& &C_1\,G_{04}^{10}\Bigl( z\,|1/2,\,1,\,-1/2,\,-1\Bigr)\,+\label{solutiong}\\
& &C_2\,G_{04}^{10}\Bigl( z\,|1,\,1/2,\,-1/2,\,-1\Bigr)\,.
\nonumber
\end{eqnarray}
We have also conditions
\begin{eqnarray}
& &1 + 8\int_0^{z_0} F(z) dz =
\frac{87 g \sqrt{N}}{32 \pi}\,\int_0^{z_0}F_0(z) \frac{dz}{\sqrt{z}}\,;\label{g}\\
& &F(z_0)\,=\,0\,;\label{pht1}
\end{eqnarray}
and boundary conditions analogous to~(\ref{bc}). The last
condition~(\ref{pht1}) means smooth transition from the non-trivial
solution to trivial one $G\,=\,0$. Knowing form~(\ref{solutiong}) of
a solution we calculate both sides of relation~(\ref{eqFg}) in two
different points in interval $0\,<\,z\,<\,z_0$ and having four
equations for four parameters solve the set. With $N\,=\,2$ we obtain
the following solution, which we use to describe the electro-weak case
\begin{eqnarray}
& &g(z_0)\,=\,0.60366\,;\quad z_0\,=\,9.61750\,;\nonumber\\
& &C_1\,=\,-\,0.035096\,; \quad C_2\,=\,-\,0.051104\,.\label{gY}
\end{eqnarray}
We would draw attention to the fixed value of parameter $z_0$. The solution
exists only for this value~(\ref{gY}) and it plays the role of eigenvalue.
As a matter of fact from the beginning the existence of such eigenvalue is
by no means evident.

Note that there is also solution with a smaller value of $z_0=0.0095531$ and rather large
 $g(z_0)=3.1867$, which with $N = 3$
presumably corresponds to strong interaction. This solution
is similar to that considered in work~\cite{Arb07} and it will be studied elsewhere.

We have one-loop expression for electro-weak coupling $\alpha_{ew}(p^2)$
\begin{equation}
\alpha_{ew}(x)\,=\,\frac{6\,\pi\,\alpha_{ew}(x_0)}{6\,\pi\,+\,5\,\alpha_{ew}(x_0)
\ln(x/x_0)}\,; \quad x=p^2\,; \label{al1}
\end{equation}
We normalize the running coupling
by condition
\begin{equation}
\alpha_{ew}(x_0)\,=\,\frac{g(Y)^2}{4\,\pi}\,=\,0.0290;\label{alphan}
\end{equation}
where
coupling constant $g$ entering in expression
~(\ref{g}) is just corresponding to this normalization point. Note that value~(\ref{alphan}) is not far from physical
value $\alpha_{ew}(M_W)\,=\,0.0337$. To compare the two values
properly one needs a relation connecting $G$ and $M_W$, which will be obtained later on.

\section{Four-fermion interaction of heavy quarks}

Let us remind that the adequate description of low-momenta region in QCD can be achieved by an introduction of the effective Nambu -- Jona-Lasinio interaction~\cite{NJL1, NJL2} (see recent review~\cite{RV}). The spontaneous generation of NJL-type interaction was demonstrated in works~\cite{Arb05, AVZ}. In these works pions are described as bound states of light quarks, which are formed due to the effective NJL interaction with account of QCD corrections. In the present work we  explore the analogous considerations and assume that scalar fields which substitute elementary Higgs fields are formed by bound states of heavy quarks $t,\,b$. As usually we introduce
left doublet $\Psi_L\,=\,(1+\gamma_5)/2 \cdot (t,\,b)$ and
right singlet $T_R\,=\,(1-\gamma_5)/2 \cdot t$.

Then we study a possibility of spontaneous generation~\cite{Arb04, Arb05, AVZ, AVZ2}
of the following effective non-local four-fermion
interaction
\begin{eqnarray}
& &L_{ff} = G_1 \bar \Psi^{\alpha}_L T_{R\,\alpha}\,\bar T_R^{\beta} \Psi_{L\,\beta}+ \nonumber\\
& &G_2 \bar \Psi^{\alpha}_L T_{R\,\beta} \bar T_R^{\beta} \Psi_{L\,\alpha}\,+\nonumber\\
& & \frac{G_3}{2} \bar \Psi^{\alpha}_L \gamma_\mu \Psi_{L\,\alpha} \bar \Psi_L^{\beta} \gamma_\mu \Psi_{L\,\beta}+ \label{ff}\\
&& \frac{G_4}{2} \bar T^{\alpha}_R \gamma_\mu T_{R\,\alpha} \bar T_R^{\beta} \gamma_\mu T_{R\,\beta}\,.\nonumber\\
& &\nonumber
\end{eqnarray}
where $\alpha,\,\beta$ are color indices. We shall formulate and solve compensation equations for form-factors of the first two interaction, while consideration of the two last ones is postponed for the next approximations. Here we follow the procedure used in works~\cite{AVZ,  AVZ2}, which deal with
four-fermion Nambu--Jona-Lasinio interaction. However coupling constants $G_3,\,G_4$ essentially influence the forthcoming results.

Following our method (see details in~\cite{AVZ,  AVZ2}) we
come to the following compensation equations for form-factors
$F_1(x)$ and $F_2(x), \,x=p^2,\, y=q^2$, corresponding respectively to the first two terms in~(\ref{ff}). In diagram form the equation is shown at Fig. 2.
\begin{widetext}
\begin{eqnarray}
& &\Phi(x)\,=\,\frac{\Lambda^2_a(N^2 G_1^2+2 N G_1 G_2+G_2^2)}{8 \pi^2(N G_1+G_2)}\Biggl(1-\frac{N G_1+G_2}{8 \pi^2}\int_0^{\bar Y} \Phi(y) dy\Biggr)+\Biggl(\Lambda^2_a+\frac{x}{2} \ln \frac{x}{\Lambda^2_a}-\frac{3 x}{4}\Biggr)\times\nonumber\\
& &\frac{G_1^2+G_2^2+2 N G_1 G_2+2 \bar G(N+1)(G_1+G_2)}{32 \pi^2 (N G_1+G_2)}-\frac{G_1^2+G_2^2+2 N G_1 G_2+2 \bar G(N+1)(G_1+G_2)}{2^9 \pi^4}\,K\times \Phi\,;\nonumber\\
& &F_2(x)\,=\,\frac{\Lambda^2_a G_2}{8 \pi^2}\Biggl(1-\frac{G_2}{8 \pi^2}\int_0^{\bar Y} F_2(y) dy\Biggr)+\Biggl(\Lambda^2_a+\frac{x}{2} \ln \frac{x}{\Lambda^2_a}-\frac{3 x}{4}\Biggr)\frac{G_1^2+G_2^2+2 \bar G(G_1+G_2(N+1))}{32 \pi^2 G_2}-\label{PhiF}\\
& &\frac{G_1^2+G_2^2+2 \bar G(G_1+G_2(N+1)}{2^9 \pi^4}\,K\times F_2\,;\;\Phi(x)\,=\,\frac{N G_1 F_1(x)+G_2 F_2(x)}{N G_1+G_2}\,;
\; \bar G = \frac{G_3+G_4}{2}\,;\;\Phi({\bar Y})=F_2({\bar Y})=0\,. \nonumber\\
& &K\times F_i = (\Lambda^2_a-x \ln \Lambda^2_a)\int_0^{\bar Y} F_i(y) dy-\ln \Lambda^2_a \int_0^{\bar Y} F_i(y) y dy+\frac{1}{6 x}\int_0^x F_i(y) y^2 dy+\ln x \int_0^x F_i(y) y dy +\nonumber\\
& & x\,\Bigl(\ln x - \frac{3}{2}\Bigr) \int_0^x F_i(y) dy+\int_x^{\bar Y} y(\ln y-\frac{3}{2})F_i(y) dy + x \int_x^{\bar Y} \ln y F_i(y) dy +
\frac{x^2}{6}\int_x^{\bar Y} \frac{F_i(y)}{y} dy\,.\label{kernel}
\end{eqnarray}
\end{widetext}
Here $N=3$, $\bar Y$ is the upper limit of integration analogous to that used in the previous section 
and $\Lambda_a$ is an auxiliary cut-off, which is introduced for loop diagrams of Fig. 2 without form-factors. It is important to stress that $\Lambda_a$ disappears from all final expressions provided all conditions for solutions being fulfilled.

Introducing substitution $G_1=\rho\,\bar G,\,G_2=\omega \bar G$ and comparing the two equations~(\ref{PhiF}) we get convinced, that both equations become being the same under the following condition
\begin{equation}
\rho=0\,.\label{x}
\end{equation}
With definitions
\begin{equation}
z\,=\,\frac{(\omega^2+8 \omega)\bar G^2 x^2}{2^{14}\,\pi^4}\,;\quad  t\,=\,\frac{(\omega^2+8 \omega)\bar G^2 y^2}{2^{14}\,\pi^4}\,;
\end{equation} 
we are rested with  one equation
\begin{widetext}
\begin{eqnarray}
& &F_2(z) = \frac{\sqrt{\omega^2+8 \omega}}{\omega }\sqrt{z}(\ln z-3)-8 \Biggl[\frac{1}{3\sqrt{z}}\int_0^z F_2(t)
\sqrt{t} dt+\sqrt{z}\,(\ln z-3)\int_0^z\frac{F_2(t)}{\sqrt{t}}dt +
\ln z\int_0^z F_2(t)\,dt\,+\,\nonumber\\
& &
\int_z^{\bar z_0} (\ln z -3) F_2(t)\,dt\,+\sqrt {z}\int_z^{\bar z_0} \ln z \frac{F_2(t)}{\sqrt{t}}\,dt\,+\,\frac{z}{3}\int_z^{\bar z_0}\frac{F_2(t)}{t}\,dt
\Biggr]\,;\quad \bar z_0\,=\,\frac{(\omega^2+8 \omega)\bar G^2 \bar Y^2}{2^{14}\,\pi^4}\,.\label{eqz}
\end{eqnarray}
\end{widetext}
Here we omit all terms containing auxiliary cut-off $\Lambda$
due to their cancellation.

Performing consecutive differentiations of Eq.(\ref{eqz}) we
obtain the following differential equation for $F_2$
\begin{eqnarray}
& &\Biggl(z\frac{d}{dz}+\frac{1}{2}\Biggr)\Biggl(z\frac{d}{dz}\Biggr)\Biggl(z\frac{d}{dz}\Biggr)\Biggl(z\frac{d}{dz}-\frac{1}{2}\Biggr)\times\label{diffeq}\\
& &\Biggl(z\frac{d}{dz}-\frac{1}{2}\Biggr)\Biggl(z\frac{d}{dz}-1\Biggr)\,F_2(z)+\,z\,F_2(z)\,=\,0\,;
\nonumber
\end{eqnarray}
Equation~(\ref{diffeq}) is equivalent to integral equation~(\ref{eqz}) provided the following boundary conditions being
fulfilled
\begin{eqnarray}
& &\int_0^{\bar z_0}\frac{F_2(t)}{\sqrt{t}}dt\,=\frac{\sqrt{\omega^2+8 \omega}}{8\,\omega }\,;\quad F_2(\bar z_0)\,=\,0\,;\nonumber\\
& &\int_0^{\bar z_0} F_2(t)
\sqrt{t}\,dt\,=\,0\,;\quad
\int_0^{\bar z_0} F_2(t)\,dt\,=\,0\,.\label{bc1}
\end{eqnarray}
Note that just boundary conditions~(\ref{bc1}) lead to cancellation of all terms containing $\Lambda_a$.
Differential equation~(\ref{diffeq}) is a Meijer equation~\cite{be} and the solution of the problem~(\ref{diffeq}, \ref{bc1}) is the following 
\begin{equation}
F_2(z)\,=\,\frac{1}{2 \sqrt{\pi}} G^{40}_{06}\Bigl(z|0,\frac{1}{2},\frac{1}{2},1,-\frac{1}{2},0\Bigr);\quad z_0 = \infty.\label{sol22}
\end{equation}
Here we also take into account condition $F_2(0)=1$ that gives
\begin{equation}
\omega=\frac{8}{3}\,.\label{omega}
\end{equation}
We would draw attention to the fact, that unique solution~(\ref{sol22}) exists only for infinite upper limit in integrals.
\section{Doublet bound state $\bar \Psi_L\,T_R$ }

Let us study a possibility of spin-zero doublet bound state $\bar \Psi_L\,T_R\,=\,\phi$, which can be referred to a Higgs scalar doublet.
With account of interaction~(\ref{ff}) using results of the previous section we have the following Bethe--Salpeter equation, in which we take into account the $t$-quark mass (see Fig. 3 without the two last diagrams)
\begin{equation}
\Psi(x)\,=\,\frac{\bar G_2}{16 \pi^2}\int \Psi(y)\,dy\,+\,
\frac{ G^2_2}{2^7\pi^4}\,K^*\times \Psi\,;\label{BS}
\end{equation}
where the modified integral operator $K^*$ is defined in the same way as operator~(\ref{kernel}) with $\bar Y\,=\,\infty$ and lower limit of integration $0$ being changed for $m^2_t$,
where $m_t$ is  the $t$-quark mass (see~\cite{AVZ}).

Then we have again differential equation
\begin{eqnarray}
& &\Biggl(z\frac{d}{dz}-a_1\Biggr)\Biggl(z\frac{d}{dz}-a_2\Biggr)\Biggl(z\frac{d}{dz}\Biggr)\Biggl(z\frac{d}{dz}-\frac{1}{2}\Biggr)\times\nonumber\\
& &\Biggl(z\frac{d}{dz}-\frac{1}{2}\Biggr)\Biggl(z\frac{d}{dz}-1\Biggr)\,\Psi(z)-
z\,\Psi(z)\,=\,0\,;\label{diffBS}\\
& &a_1\,=\,-\,\frac{1+\sqrt{1+64\,\mu}}{4}\,;\; a_2\,=\,-\,\frac{1-\sqrt{1+64\,\mu}}{4}\,;\nonumber\\
& &\mu=\frac{G_2^2\,m^4_t}{2^{12} \,\pi^4}\,.\nonumber
\end{eqnarray}
where the main difference is the other sign of the last term,
while variable $z$ is just the same as in~(\ref{diffeq}) with account of relation~(\ref{omega}).
Boundary conditions now are the following
\begin{eqnarray}
& &\int_\mu^\infty\frac{\Psi(t)}{\sqrt{t}}dt\,=\,0;\;
\int_\mu^\infty \Psi(t)
\sqrt{t}\,dt\,=\,0;\nonumber\\
& &\int_\mu^\infty \Psi(t)\,dt\,=\,0;\; \;\Psi(\mu)\,=\,1.\label{bcBS}
\end{eqnarray}
Solution of the problem is presented in the following form

\begin{eqnarray}
& &\Psi(z)\,=\,\bar C_1\,G^{50}_{06}(z|a_1,a_2,1/2,1/2,1,0)+\nonumber\\
& &\bar C_2\,G^{30}_{06}(z|0,1/2,1,a_1,a_2,1/2)+\label{Psi}\\
& &\bar C_3\,G^{30}_{06}(z|1/2,1/2,1,a_1,a_2,0)+\nonumber\\
& &\bar C_4\,G^{50}_{06}(z|a_1,a_2,0,1/2,1,1/2)\,;\nonumber
\end{eqnarray}
where $\bar C_i$ for given $\mu$ are uniquely defined by conditions~(\ref{bcBS}).

We define interaction of the doublet $\phi$ with heavy quarks
\begin{equation}
L_\phi\,=\,g_\phi(\phi^*\bar \Psi_L\,T_R\,+\,\phi\,\bar T_R\,\Psi_L)\,;
\end{equation}
where $g_\phi$ is the coupling constant of the new interaction to be defined by normalization condition of
the solution of equation~(\ref{BS}). Then we take into
account the contribution of interaction of quarks with
gluons and the exchange of $\phi$ as well (see Fig. 3).
Using standard perturbative method we obtain for mass
of the bound state under consideration the following expression in the same way as in~\cite{AVZ}.
\begin{eqnarray}
& &m_\phi^2\,=\,-\,\frac{m^2_t\,I_5}{\pi\,\sqrt{\mu}\,I_2}\,;\quad I_2\,=\,\int_\mu^\infty\frac{\Psi(z)^2\,dz}{z}\,;\label{mphi}\\
& &I_5\,=\,\int_\mu^\infty\frac{(16\,\pi\,\alpha_s(z)-\,g_\phi^2)\,\Psi(z)\,dz}{16\,\pi\,z}\int_\mu^z\frac{\Psi(t)\,dt}{\sqrt{t}}\,.\nonumber
\end{eqnarray}
Here $\alpha_s(z)$ is the strong coupling with standard evolution, normalized at the $t$-quark mass.
Provided term with brackets inside $I_5$ being positive, bound state $\phi$ is a tachyon.
Let us recall the well-known relation for $t$-quark mass, which is defined by non-zero vacuum average $\eta$ of $(\phi_2^*+\phi_2)/\sqrt{2}
$. It reads
\begin{equation}
m_t\,=\,\frac{g_\phi\,\eta}{\sqrt{2}}\,;\label{tusual}
\end{equation}
where $\eta=246.2\,GeV$ is the value of the electro-weak scalar condensate. However in our approach there are additional contribution to this mass, e.g. due to diagram shown at Fig. 4. For the moment we can not calculate contributions of such diagrams, so we take for experimental value of the $t$-quark mass the modified definition
\begin{equation}
m_t\,=\,\frac{g_\phi\,\eta}{\sqrt{2}\,f}\,;\label{tour}
\end{equation}
where we take into account additional contributions in terms
of factor $f$. This means that $m_t$ will be also the input parameter. 

Let us consider the case when relation~(\ref{mphi}) leads to a tachyon state. For Higgs mechanism to be realized we need also four-fold interaction
\begin{equation}
\L_{\phi4}\,=\,\lambda\,(\phi^*\phi)^2\,.\label{4phi}
\end{equation}
Coupling constant in~(\ref{4phi}) is defined in terms of
the following loop integral
\begin{equation}
\lambda\,=\,\frac{3\,g_\phi^4}{16\, \pi^2}\,I_4\,;\quad I_4\,=\,\int_\mu^\infty
\frac{\Psi(z)^4\,dz}{z}\,.\label{lambda}
\end{equation}
From well-known relations $\eta^2\,=\,-m_\phi^2/\lambda$ and  the Higgs mass squared $M_H^2\,=\,-\,2\,m_\phi^2$ we have
\begin{equation}
\eta^2\,=\,\frac{16 \pi\,m^2_t\,I_5}{3\,g_\phi^4\,\sqrt{\mu}\,I_2\,I_4}\,; \quad M_H^2\,=\,\frac{2\,m^2_t\,I_5}{\pi\,\sqrt{\mu}\,I_2}\,.\label{eta2}
\end{equation}
From~(\ref{tour}) and~(\ref{eta2}) we have useful relation
\begin{equation}
2\,=\,\frac{16 \pi\,I_5}{3\,g_\phi^2\,f^2\,\sqrt{\mu}\,I_2\,I_4}\,.\label{2}
\end{equation}
We define $g_\phi$ from a normalization condition, which we formulate in the next section.\\
\\

\section{Interaction of the Higgs field with bosons W}
Let us consider an interaction of $W$-bosons and the Higgs particle. We start with the ready
phenomenology according to which the $W\,W\,H$ interaction
corresponds to the following vertex
\begin{equation}
\imath\,g_{\mu \nu}\,\delta_{a b}\,g\, M_W\,.\label{wwh0}
\end{equation}
From this moment we assume that $M_W$ is known.
Additional contribution to vertex~(\ref{wwh0}) is provided by our effective
interaction due to diagram presented at Fig. 5. Substituting into this diagram the
first approximation $F_0$~(\ref{solution}, \ref{z0C}, \ref{norm}) for the three-boson form-factor we
obtain term with Lorentz structure
$$
(g_{\mu\nu}\,(q\,k)\,-\,q_\nu\,k_\mu)\,;
$$
and thus we are to look for 
the following additional gauge invariant contribution to 
$W\,W\,H$ vertex
\begin{eqnarray}
& & \imath\,\bar H\,\delta_{a b}\,(g_{\mu
\nu}\,(q\,k)\,-\,q_\nu\,k_\mu)\,F_H(x)\,;\label{wwh1}\\
& &\,\nonumber\\
& & \bar H\,=\,\frac{\sqrt{2}\,G\,g\,M_W\,A\,}{\pi};\quad F_H(0)=1\,;\nonumber
\end{eqnarray}
where $F_H(x)$ is a form-factor, which we find by solving 
an equation corresponding to diagrams at Fig. 6 and $A$ is a constant to be defined from the solution of this equation. Following the same approximation as in Section 2 we also take into account terms being proportional to gauge coupling $g$ and obtain the following equation (the upper limit of the momentum integration is the same $Y$ as in Eq.(\ref{eqF}))
\begin{widetext}
\begin{eqnarray}
& & F_H(t)\,=\,INH(t) - \frac{2}{3 t}\int_0^t F_H(t') t'dt'+ \frac{4}{3 \sqrt{t}}\int_0^t F_H(t') \sqrt{t'}dt'+ \frac{4\sqrt{t}}{3 }\int_t^{z_{01}} \frac{F_H(t')}{ \sqrt{t'}}dt'-\frac{2 t}{3 }\int_t^{z_{01}} \frac{F_H(t')}{t'}dt'+\nonumber\\
& & \frac{\xi}{4 \pi}\biggl(\frac{1}{6 t}\int_0^t F_H(t') t'dt'+ \frac{1}{2 \sqrt{t}}\int_0^t F_H(t') \sqrt{t'}dt' + \frac{1}{2 }\int_t^{z_{01}} \frac{F_H(t')}{ \sqrt{t'}}dt'+\frac{\sqrt{t}}{6 }\int_t^{z_{01}} \frac{F_H(t')}{t'}dt'\biggr)\,.\label{EQH}\\
& &\xi=\frac{\bar H^2}{G}\,;\quad t=\frac{G^2\,x^2}{64\,\pi^2}\,;\quad z_{01}\,=\,8\,z_0\,;\quad F_H(z_{10})\,=\,0\,.\nonumber
\end{eqnarray}
\end{widetext}
We would draw attention to important parameter $\xi$, which defines a ratio of coupling constant squared of interaction~(\ref{wwh1}) and the main three boson coupling~(\ref{FFF}). 
We shall see that equation~(\ref{EQH}) has solutions only for 
some definite values of this parameter and so we have a set of eigenvalues of $\xi$.  
The inhomogeneous part of Eq.(\ref{EQH}) reads
\begin{widetext}
\begin{eqnarray}
& & INH(t)\,=\,\frac{1}{A}\Biggl(\int_0^{z_0}\frac{F^2(z)}{\sqrt(z)}\,dz\,+\,\sqrt{\frac{t}{8}}\biggl(-\,\frac{31}{18}+\frac{1}{3}\int_0^{z_0}\frac{(F(z)-1)^2}{z}\,dz+\,\frac{1}{3}\ln\frac{8\,z_0}{t}\biggr)\Biggr)\,+\,4\int_0^{z_{01}}F_H(t')dt'\,+\nonumber\\
& &\,\frac{g}{\pi}\Biggl(\frac{1}{t}\int_0^t F_{TH}(t')\sqrt{t'}dt'-\frac{3}{\sqrt{t}}\int_0^t F_{TH}(t') dt' - 3\int_t^{z_0}\frac{F_{TH}(t')}{\sqrt{t'}} dt' + \sqrt{t}\int_t^{z_0}\frac{F_{TH}(t')}{t'} dt'+\,F_H(t)\times\label{INH}\\
& &\biggl(-\,\frac{5}{2 t}\int_t^{z_0}\frac{F_T(t')}{\sqrt{t'}} dt'\,+\,\frac{7}{\sqrt{t}}\int_0^t F_T(t') dt'\,-\,3\int_0^{t}\frac{F_T(t')}{\sqrt{t'}} dt'\,+\,3\int_t^{z_0}\frac{F_T(t')}{\sqrt{t'}} dt'-\sqrt{t}\int_t^{z_0}\frac{F_T(t')}{t'} dt'\,-\nonumber\\
& &\,\frac{t}{2}\int_t^{z_0}\frac{F_T(t')}{t'\sqrt{t'}} dt'\biggr)\Biggr)\,;\quad t=\frac{G^2\,x^2}{64\,\pi^2}\,;\quad F_T(t)=F\Bigl(\frac{t}{8}\Bigr)\,;\quad F_{TH}(t)=F_T(t)F_H(t)\,;\nonumber
\end{eqnarray}
\end{widetext}
Using already obtained form-factor $F(z)$~(\ref{solutiong})  and values of parameters $g,\,z_0$~(\ref{gY}) we solve equation~(\ref{EQH}) numerically. Let us define parameter $\Lambda$ in the following way
\begin{equation}
G\,=\,\frac{\Lambda}{M_W^2}\,.\label{Lambda}
\end{equation}
Vertex~(\ref{wwh1}) enters also to normalization condition
defining coupling $g_\phi$ according to diagrams shown at Fig. 7, where symbol of diagram denotes its coefficient afore $p^2$ in an expansion in $(p^2)^n$ with $p$ being a momentum of the field $\phi$.
 It reads
\begin{eqnarray}
& &\frac{3 g^2_\phi}{32 \pi^2}\,\Biggl(I_2\,+\,\frac{\alpha_s}{4 \pi}\Bigl(I_{22}^2+2\,I_6\Bigr)\Biggr)\,+\,I\,=\,1;\label{normgh}\\
& &I\,=\,I_{hh}\,+\,I_{gh}\,;\quad I_{hh}\,=\,\frac{H^2}{16 \pi^2}\int_{\mu'}^{z_{01}}\frac{F_H^2\,dt}{\sqrt{t}}\,;\nonumber\\
& &I_{gh} = \frac{g H}{16 \pi^2}\int_{\mu'}^{z_{01}}\frac{F_H dt}{\sqrt{t}}\,;\;I_{22} = \int_\mu^\infty\frac{\Psi(t) dt}{t}\,;\nonumber\\
& &
I_6 = \int_\mu^\infty\frac{\Psi(z) dz}{z\sqrt{z}}\int_\mu^z\frac{\Psi(t) dt}{\sqrt{t}}\,;\; \mu'\,=\,\frac{\Lambda^2}{64\,\pi^2}.\nonumber
\end{eqnarray}
There are several solutions of equation~(\ref{EQH}) with different eigenvalues of $\xi$ giving the following values of parameters:
\begin{eqnarray}
& &\xi = 7.27;\,\Lambda = 0.0157;\,A=-79.1;\,I=0.165;\label{xi1}\\
& &\xi = 35.2;\,\Lambda = 0.0371;\,A = 113.4;\,I = 0.470;\label{xi2}\\
& &\xi = 69.3;\,\Lambda = 0.0199; A = -217.3; I = 0.756;\label{xi3}\\
& &\xi = 116.4;\,\Lambda = 0.0042; A = 615.2;\,I =   0.845;\label{xi4}\\
& &\xi = 162.0;\,\Lambda = 0.001; A = -1484;\,I = 0.996.\label{xi5}
\end{eqnarray}
Small values of $I$ lead to too large values of $g_\phi$ and thus to positive values of $m^2_\phi$. This means an absence of Higgs mechanism. Too large values of $I$ lead to imaginary $g_\phi$. Our analysis shows, that only two intermediate solutions~(\ref{xi3}, \ref{xi4}) are physically admissible.
Note, that values of parameter $\Lambda$ for these solutions
satisfy limitations~(\ref{lambda1}).

Other interesting parameters depend on value $\alpha_s$ of strong coupling at the $t$-quark mass. We have preferable value of this quantity, which is obtained from value $\alpha_s(M_Z)\,=\,0.1184 \pm 0.0007$. Using standard evolution of $\alpha_s$ we obtain
\begin{equation}
\alpha_s\,=\,\alpha_s(172\,GeV)\,=\,0.108\,.\label{as}
\end{equation}
However we perform calculations also for nearby values of
$\alpha_s$ in the interval from 0.08 up to 0.12. In doing this we proceed in the following way: for six parameters
$\mu,\,g_\phi,\,\eta,\,m_t,\,M_H,\,f$ we have five  relations~(\ref{al1}, \ref{mphi},\ref{tour}, \ref{eta2},  \ref{normgh}) and the well-known expression
\begin{equation}
M_W\,=\,\frac{g_w\,\eta}{2}\,;\label{mw}
\end{equation}
where $g_w$ is weak interaction constant $g$ at $W$ mass. We obtain it by usual RG evolution expression~(\ref{al1}) from value $g$ at $Y$~(\ref{gY}).
Let us remind that we consider $M_W$ as an input. All the  parameters entering the expressions excepting $f$ are already defined above.
At the present stage of the study we can not calculate $f$, because e.g. for diagrams at Fig. 4 one needs knowledge of behavior of form-factor $F_H$ in dependence of all three variables. This problem is not solved yet, therefore we define value $f$ by fixing the $t$-quark mass in relation~(\ref{tour}). We have found, that for any solution of our set of equation $M_W\,=\,80.4\,GeV$ is reproduced via relation~(\ref{mw}) for any value of $f$ in~(\ref{tour}). Thus for the moment we have two input parameters, which are safely known from the experiment
\begin{equation}
M_W\,=\,80.399\pm 0.023\,;\quad M_T\,=\,172.0\pm 1.6\,.\label{input}
\end{equation}

Now we present thus obtained parameters in Table 1 for the two solutions~(\ref{xi3}, \ref{xi4}) being mentioned above.
We see from the Table that important parameters $\eta$ and $g_w$ are obtained in close agreement to experimental data.  For solution~(\ref{xi3}) $\eta$ and $g_w$ both differ from the experimental numbers not more than by $4.7\%$ and for solution~(\ref{xi4}) not more than by $4.0\%$. We present at the Table values of parameters in dependence on value of
strong coupling at the $t$-quark mass. We see that this dependence is practically insignificant for all parameters but four-fermion coupling $G_2$. For final result we would choose central experimental value $\alpha_s(M_T)\,=\,0.108$. So we take as an estimate of our precision for Higgs boson mass just the precision of definition of $\eta$ and $g_w$ and thus we obtain two possible values for the Higgs boson mass:
\begin{eqnarray}
& &\xi\,=\,69.3030\,:\quad M_{H1}\,=\,(306\,\pm\,14)\,GeV\,;\label{MH1}\\
& &\xi\,=\,116.436\,:\quad M_{H2}\,=\,(253\,\pm\,10)\,GeV\,.\label{MH2}
\end{eqnarray}
This prediction is the main result of the work.  
Note that the method to calculate masses~(\ref{MH1}, \ref{MH2}) in our approach was previously checked while calculating $\pi$-meson mass due to NJL-type interaction~\cite{AVZ}. With pion decay constant $f_\pi$ being fixed we have~\cite{AVZ} $m_\pi \simeq 135 MeV$, that differs from real  charged pion mass just by few per cent.

With knowledge of parameter $\Lambda$~(\ref{xi3}, \ref{xi4}) we have thorough information for definition of form-factor of the anomalous three-boson interaction~(\ref{solutiong})
\begin{equation}
F(z)\,;\qquad z\,=\,\frac{\Lambda^2\,Q^4}{512\,\pi^2\,M_W^4}\,.\label{FQ}
\end{equation}
The behavior of $F(Q)$ for solution~(\ref{xi4}) is presented at Fig. 8.
\section{Experimental implications}
New interaction of Higgs field $H$ with $W$-s
\begin{equation}
L_{HWW}\,=\,\frac{\bar H}{2}\,W^a_{\mu \nu}\,W^a_{\mu \nu}\,H.\label{inthww1}
\end{equation}
leads to changes in usual branching ratios for $H$ decays. We use here the well-known expression for $W^0$  mixed state with physical value for $\sin^2\theta_W$. There are unusually significant channels
$H \to \gamma\,\gamma$ and $H \to \gamma\,Z$. Therefore there may be additional restrictions from existing experiments. Data from Tevatron on search for
Higgs particle in $\gamma\,\gamma$ channel~\cite{ggd0} exclude Higgs particle with
interaction~(\ref{inthww1}) for $M_H\,<\,150\,GeV$. Our predictions~(\ref{MH1}, \ref{MH2}) evidently do not contradict to this result. The recent CMS results from LHC~\cite{CMS} also do not impose restrictions for bosons with masses~(\ref{MH1}, \ref{MH2}).

Let us remind that we have obtained all the necessary parameters for estimation of widths and branching ratios
for the two variants~(\ref{MH1}, \ref{MH2}). The results are presented at Table 2, where we have taken central values of masses~(\ref{MH1}, \ref{MH2}).

We also estimate cross-sections of the Higgs boson production at LHC with $\sqrt{s}\,=\,7\,TeV$.
\begin{eqnarray}
& &M_H\,=\,306\,GeV:\quad \sigma_t\,\simeq\,26\,pb;\label{sig1}\\
& &M_H\,=\,253\,GeV:\quad \sigma_t\,\simeq\,14\,pb.\label{sig2}
\end{eqnarray}
Note that rather unexpected prediction: larger cross-section for larger Higgs mass, is easily understood by comparison of values of parameters $\Lambda$ and $A$ for these solutions (\ref{xi3}, \ref{xi4}).

For example with mass $253\,GeV$ under conditions of work~\cite{CMS} we would have 10 events in interval $240\,GeV < M_H < 270\,GeV$ for $H\to W^+\,W^-\to 2l2\nu$ provided efficiency of registration being 100\%. To compare with plot in~\cite{CMS} we have with mass of the Higgs boson indicated in brackets
\begin{eqnarray}
& &\sigma BR(H\to WW\to 2l2\nu) = 0.28\,pb\,(253 GeV);\nonumber\\
& &\sigma BR(H\to WW\to 2l2\nu) = 0.43\,pb\,(306 GeV).\label{limit}
\end{eqnarray}
that is significantly lower, than upper limit curve in~\cite{CMS}. However with increasing of integral luminosity by an order of magnitude a quest for Higgs of the discussed type would become promising, especially for decay channel $H\to \gamma \gamma$. E.g. with integral luminosity $360\,pb^{-1}$ and
Higgs mass $253\,GeV$ we would have around 170 decays $H\to \gamma \gamma$.

For calculations of this section CompHEP package~\cite{comphep} was used.

\section{Conclusion}

To conclude we would emphasize, that albeit we discuss quite
unusual effects, we do not deal with something beyond the Standard
Model. We are just in the framework of the Standard Model. What
makes difference with usual results is {\bf non-perturbative non-trivial solution} of
compensation equation. There is of course also {\bf trivial
perturbative solution}. Which of the solutions is realized is to be
defined by stability conditions.  The problem of stability is extremely complicated and needs a special extensive study. That is why we present in the work two possible solutions~(\ref{MH1}, \ref{MH2}) in view of anticipation of the forthcoming LHC results. Confirmation of prediction~(\ref{MH1}, \ref{MH2}) (see also Tables 1, 2) would mean proof of the non-perturbative nature of the real Higgs mechanism.

With the present results we would draw attention to two important achievements provided by the non-trivial non-perturbative solution.
The first one is unique determination of gauge electro-weak coupling
constant $g(M_W)$ in close agreement with experimental
value. The second result consists in calculation of the fundamental quantity $\eta$ -- vacuum average of the scalar Higgs field. At this point we would emphasize, that the
existence of a non-trivial solution itself always leads to
additional conditions for parameters of a problem under study. These two achievements strengthen the confidence in the correctness of
applicability of Bogoliubov compensation approach to the principal problems of elementary particles theory. We consider a check of prediction for Higgs boson mass~(\ref{MH1}, \ref{MH2}) and of its properties being shown at Table 2 as a decisive test of validity of the compensation approach.

\section*{Acknowledgments}
Authors express gratitude to E.E. Boos and V.I Savrin for valuable discussions.

\newpage
\begin{center}
{\bf Table captions}
\end{center}
\bigskip
\bigskip
Table 1. Results of solution of the set of equations for
vacuum average $\eta$, EW gauge constant $g$ at $M_W$, the Higgs field coupling constant with the $t$-quark $g_\phi$ and the Higgs scalar mass $M_H$, $\xi_1=69.303,\,\xi_2=116.436$  (we fix $M_W\,=\,80.4\,GeV\,;\;M_T\,=\,172.2\,GeV$).\\
\\
Table 2. Properties of the Higgs particle for two solutions.\\
\\

\newpage
\begin{center}
{\bf Figure captions}
\end{center}
\bigskip
\bigskip
Fig. 1. Diagram representation of the compensation
equation. Black spot corresponds to anomalous three-boson
vertex with a form-factor. Empty circles correspond to point-like anomalous
three-boson and four-boson vertices. Simple point corresponds to usual gauge  vertex.
Incoming momenta are denoted by the corresponding external lines.\\
\\
Fig. 2. Diagram representation of the compensation
equation for the four-fermion interaction~(\ref{PhiF}). Lines describe quarks. Simple point
corresponds to the point-like vertex and black circle corresponds to a vertex with a form-factor.\\
\\
Fig. 3. Diagram representation of the Bethe-Salpeter equation for a bound state of heavy quarks. Double line represent the bound state and dotted line describes a gluon. Black circle corresponds to BS wave function. Other notations are the same as at Fig.2. \\
\\
Fig. 4. Diagrams for new contribution to the $t$-quark mass. Dotted line represents the $t$-quark
, other notations the same as at Figs. 2 - 4.\\
\\
Fig. 5. Diagram representation of the first contribution to  $INH$~(\ref{EQH}). Here the double line correspond to the Higgs scalar, simple point is the usual vertex~(\ref{wwh0}), other notations the same as at Fig. 1.\\
\\
Fig. 6. Diagram representation of equation for $WWH$
vertex with a form-factor. The double circle with black inside describes their form-factor. The same with empty internal circle correspond to point-like anomalous  $H\,W\,W$ vertex. Single circle corresponds to $H\,W\,W$ vertex
calculated according diagram at Fig. 5.\\
\\
Fig. 7. Diagram representation of normalization condition of
Higgs scalar interaction with heavy quark. Dotted lines below represent W. Other notations the same as at Figs. 2, 3.\\
\\
Fig. 8. Behavior of form-factor $F(Q)$~(\ref{solutiong}) for the second solution~(\ref{MH2}), $0 < Q < 18.5\,TeV$. For $Q >18.495\,TeV$, $F(Q)\,=\,0$.
\newpage
\begin{widetext}
\begin{center}
{\bf Table 1.}\\
\bigskip
\bigskip


\begin{tabular}{||c|c||c|c|c|c|c|c||c||}
\hline\hline
      \multicolumn{2}{||c||} \,&\multicolumn{6}{c||}{$\alpha_s(M_T)$} &\multicolumn{1}{c||} {Experiment}\\     \cline{3-9}
                     \multicolumn{2}{||c||} \,&0.08&0.09&0.10&0.108&0.11&0.12&\multicolumn{1}{c||}{ $0.108\pm 0.001$} \\ \hline   \hline
       $M_W\,GeV$&$\xi_1$&80.4&80.4&80.4&80.4&80.4&80.4& \\ \cline{2-8}
 (input)      &$\xi_2$&80.4&80.4&80.4&80.4&80.4&80.4& $80.399\pm 0.023$\\        \hline
$M_T\,GeV$&$\xi_1$&172.2&172.2&172.2&172.2&172.2&172.2&\\ \cline{2-8}
       (input) &$\xi_2$&172.2&172.2&172.2&172.2&172.2&172.2& $172.0\pm 1.6$ \\        \hline
        $\eta\,GeV$&$\xi_1$&258,23&258.26&258.26&258.26&258.26&258.27& \\ \cline{2-8}
       &$\xi_2$&256.61&256.61&256.57&256.58&256.58&256.62&$246.22\pm 0.03$ \\        \hline
         $g_w(M_W)$&$\xi_1$&0.623&0.623 &0.623 &0.623&0.623&0.623& \\ \cline{2-8}
       $\,$&$\xi_2$&0.627&0.627&0.627&0.627&0.627&0.626& $0.651\pm 0.002$  \\        \hline
         $g_\phi$&$\xi_1$&1.400&1.421&1.438&1.450&1.453&1.466& \\ \cline{2-8}
       $\,$&$\xi_2$&1.324&1.344&1.361&1.374&1.377&1.391& $\,?$\\        \hline
        $M_H\,GeV$&$\xi_1$&303.50&304.68&305.35&305.63&305.67&305.74& \\  \cline{2-8}
       &$\xi_2$&250.29&251.51&252.34&252.87&252.95&253.38& $\,?$\\   \hline
        $G_2\,M_T^2$&$\xi_1$&0.1314&0.1658&0.2022&0.2329&0.2405&0.2803& \\  \cline{2-8}
       &$\xi_2$&0.2493&0.3172&0.3908&0.4536&0.4699&0.5544& $\,?$\\ \hline
       \end{tabular}


\qquad\qquad
\end{center}
\end{widetext}

\newpage

\begin{widetext}
\begin{center}
{\bf Table 2.}\\
\bigskip
\large{
\begin{tabular}{|c|c|c|c|c|c|}
\hline
$M_H\,GeV$&$\Gamma_H\,GeV$ & $H\to W^+W^-$ & $H\to ZZ$ & $H\to Z\gamma$ & $H\to \gamma \gamma$  \\
\hline
306 & 101 &   41.0\% & 27.8\% &26.2\% & 5.0\% \\
\hline
253 & 29.9  & 50.3\% & 31.2\% & 15.2\% & 3.3\%  \\
\hline

\end{tabular}
}
\end{center}
\end{widetext}
\onecolumngrid
\newpage
\includegraphics{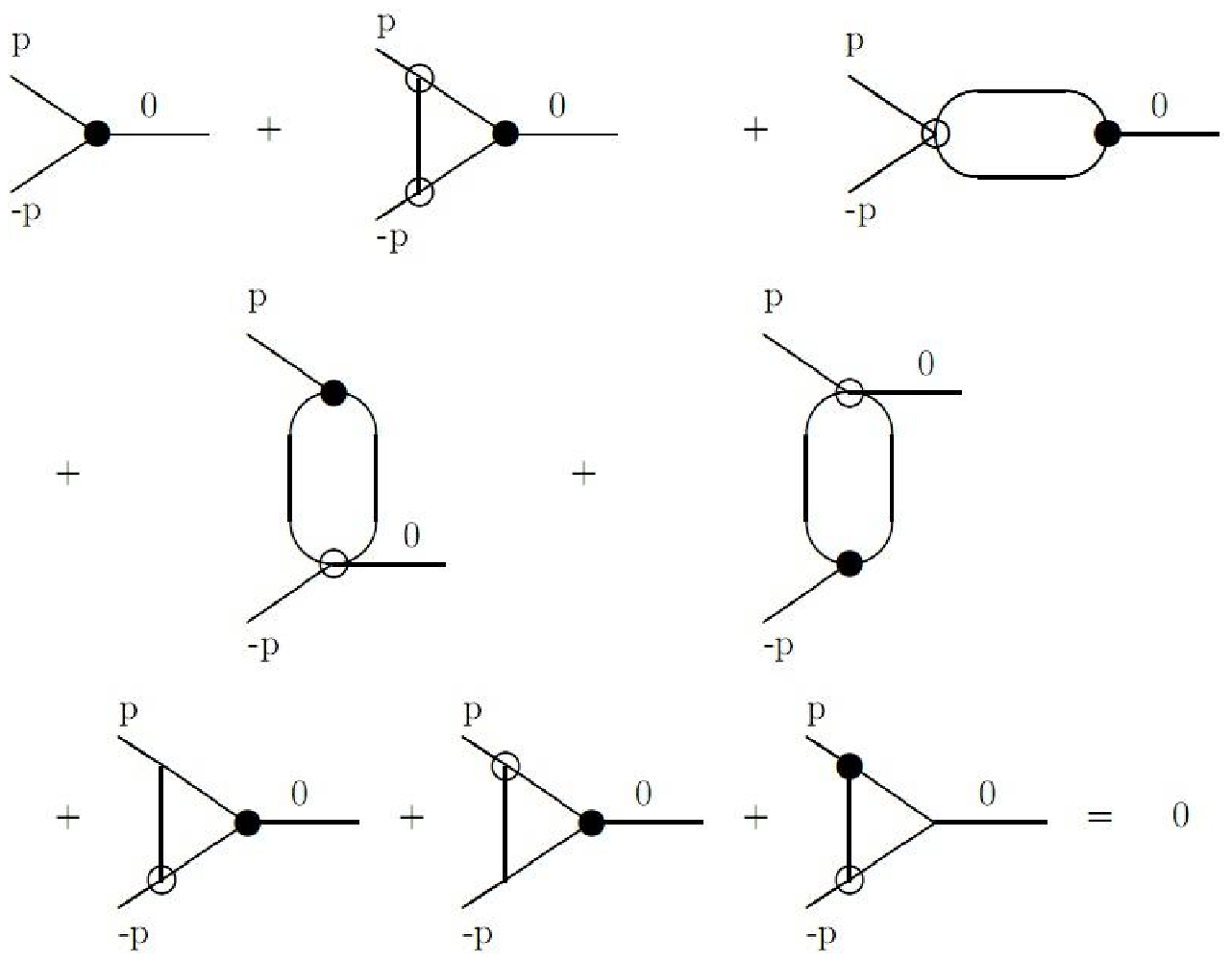}
\begin{center}
Fig. 1.
\end{center}

\newpage
\newpage
\includegraphics{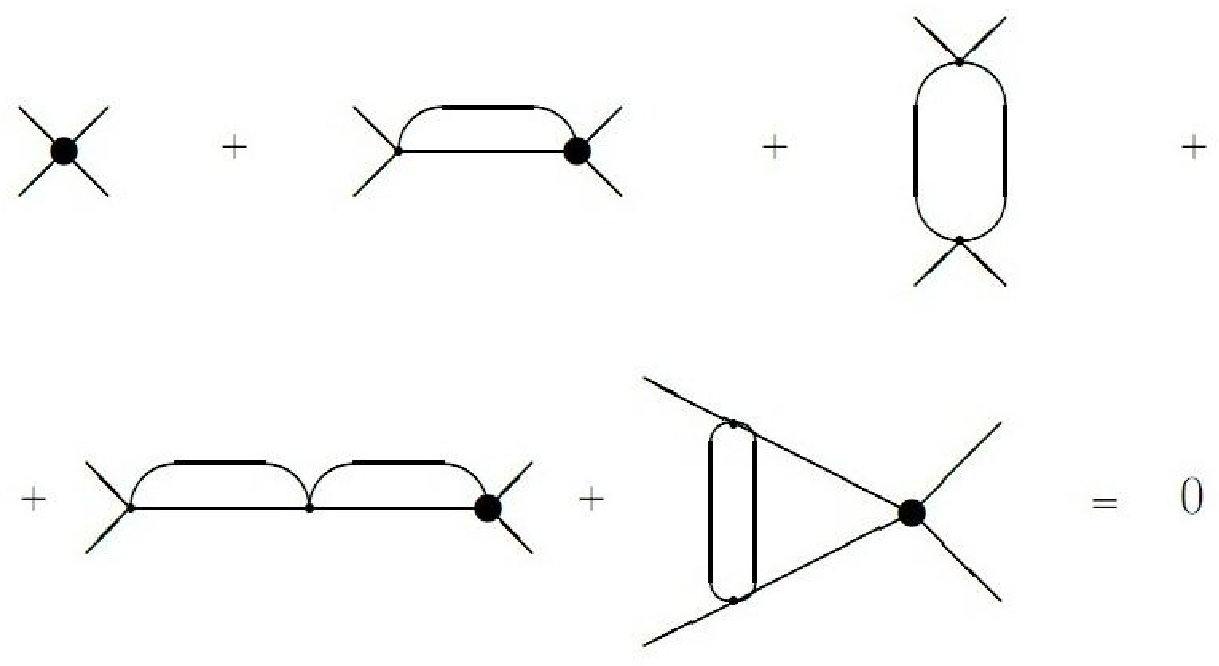}\begin{center}
Fig. 2.
\end{center}
\newpage
\newpage
\includegraphics{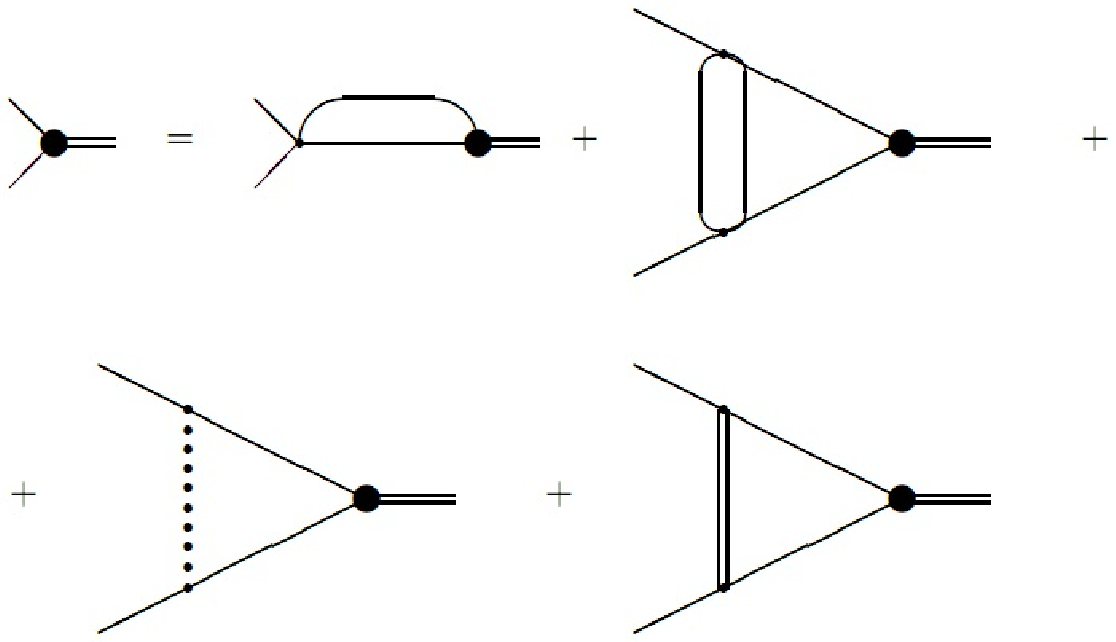}\begin{center}
Fig. 3.
\end{center}
\newpage
\newpage
\includegraphics{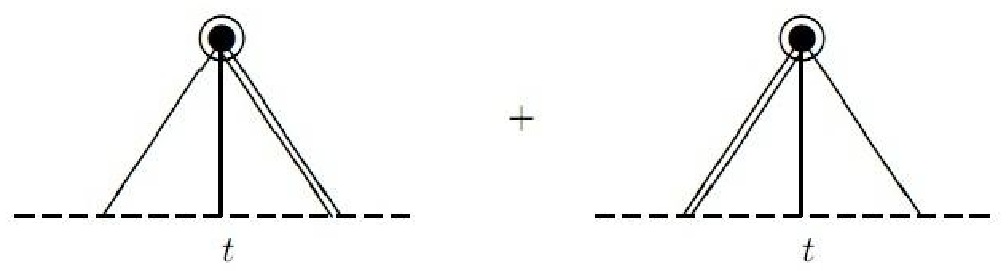}(10,10)\begin{center}
Fig. 4.
\end{center}
\newpage
\newpage
\begin{center}
\includegraphics{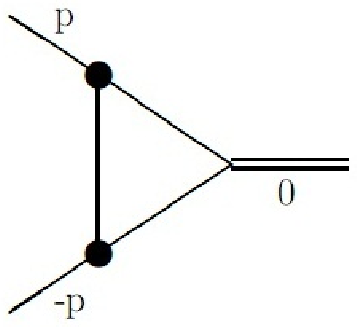}
Fig. 5.
\end{center}
\newpage
\newpage
\includegraphics{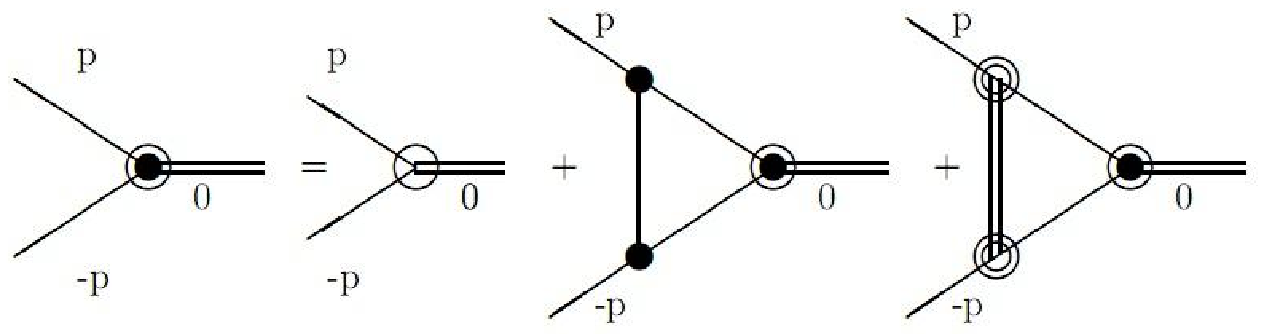}\begin{center}
Fig. 6.
\end{center}
\newpage
\includegraphics{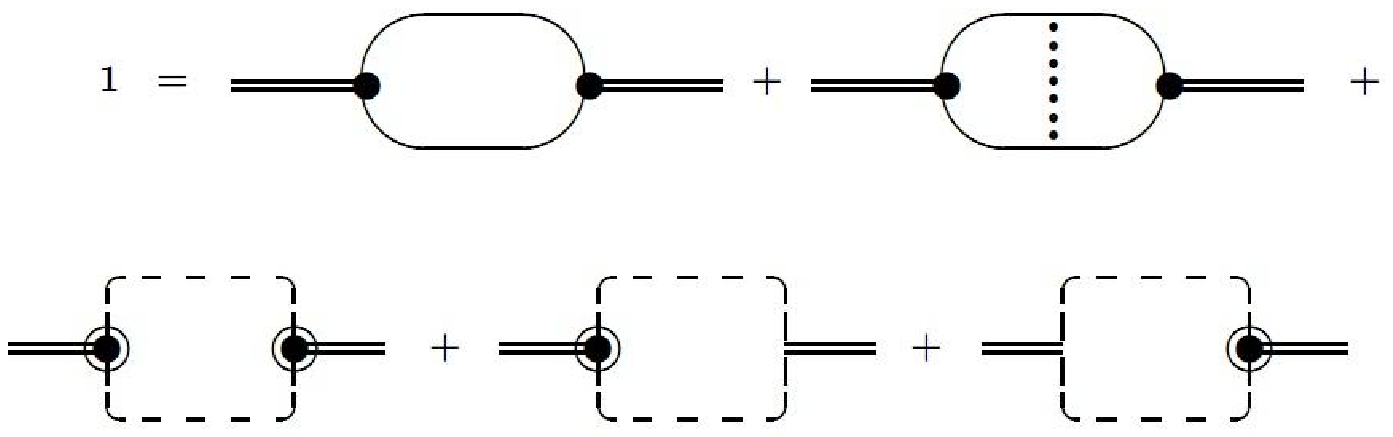}
\begin{center}
Fig. 7.
\end{center}
\newpage
\begin{center}
\includegraphics{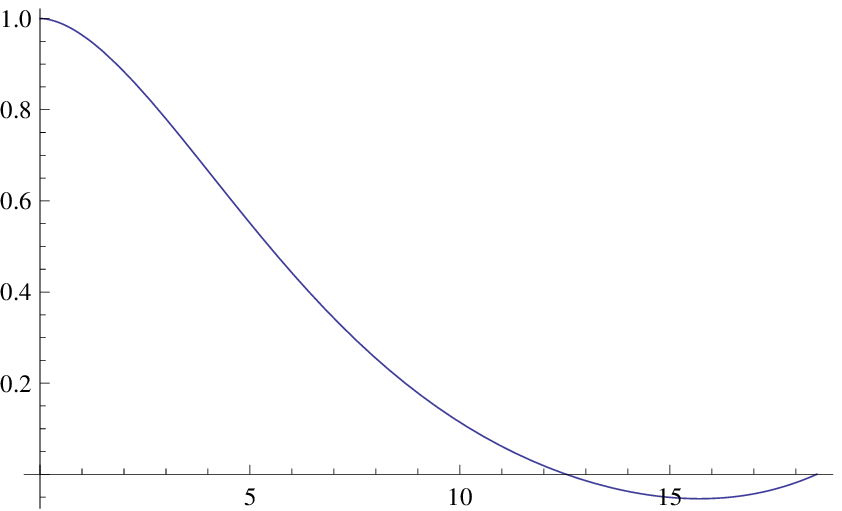}
\begin{picture}(100,150)
\put(-140,-40){Fig. 8}
\put(-265,130){$F(Q)$}
\put(-5,-2) {$TeV$}
\put(-150,-10){$Q$}
\end{picture}


\label{}

\end{center}

\begin{thebibliography}{**}
\bibitem{Arb04} B.A. Arbuzov, Theor. Math. Phys., {\bf 140}, 1205 (2004).
\bibitem{Arb05} B.A. Arbuzov, Phys. Atom. Nucl., {\bf 69}, 1588 (2006).
\bibitem{AVZ} B.A. Arbuzov, M.K. Volkov and I.V. Zaitsev, Int. Journ. Mod.
Phys. A, {\bf 21}, 5721 (2006).
\bibitem{Arb07} B.A. Arbuzov, Phys. Lett. B, {\bf 656}, 67 (2007).
\bibitem{AVZ2} B.A. Arbuzov, M.K. Volkov and I.V. Zaitsev, Int. Journ. Mod.
Phys. A {\bf 24}, 2415 (2009).
\bibitem{Arb09} B.A. Arbuzov, Eur. Phys. J. C, {\bf 61}, 51 (2009).
\bibitem{Bog1} N.N. Bogoliubov. Soviet Phys.-Uspekhi, {\bf 67}, 236 (1959).
\bibitem{Bog2} N.N. Bogoliubov. Physica Suppl., {\bf 26}, 1 (1960).
\bibitem{Bog} N.N. Bogoliubov, {\it Quasi-averages in problems of
statistical mechanics.} Preprint JINR D-781, (JINR, Dubna 1961).
\bibitem{Hag} K. Hagiwara, R.D. Peccei, D. Zeppenfeld and K. Hikasa, Nucl. Phys. B, {\bf 282}, 253 (1987).
\bibitem{Arb92} B.A. Arbuzov, Phys. Lett. B, {\bf 288}, 179 (1992).
\bibitem{EW} LEP Electro-weak Working Group, arXiv:
hep-ex/0612034v2 (2006).
\bibitem{be} H. Bateman and A. Erd\'elyi, {\it Higher
transcendental functions. V. 1} (New York, Toronto, London: McGraw-Hill,
1953).
\bibitem{NJL1} Y. Nambu and G. Jona-Lasinio, Phys. Rev. {\bf 122}, 345 (1961).
\bibitem{NJL2} Y. Nambu and G. Jona-Lasinio, Phys. Rev. {\bf 124}, 246 (1961).
\bibitem{RV} M.K. Volkov and A. Radzhabov, Phys. Usp. {\bf 49}, 551 (2006).
\bibitem{ggd0} V.M. Abazov et al. (D0 Collaboration), Phys. Rev. Lett., {\bf 102}: 231801, 2009; arXiv: 0901.1887 [hep-ex] (2009).
\bibitem{CMS} The CMS Collaboration, arXiv:1102.5429 [hep-ex] (2011).
\bibitem{comphep} E.E. Boos {\it et al.} (CompHEP Collaboration), Nucl. Instr. Meth. A {\bf 534}, 250 (2004).
\end{thebibliography}
\end{document}